\documentclass[twocolumn,prd,aps,10pt]{revtex4-1}

\usepackage[english]{babel}
\usepackage[utf8]{inputenc}
\usepackage{amssymb}
\usepackage{amsmath}
\usepackage{color}
\usepackage{hyperref}
\usepackage{bbm}
\usepackage{epsfig}

\newcommand{\e}{\varepsilon}

\newcommand{\s}{\sigma}
\newcommand{\w}{\omega}

\newcommand{\ex}{\Delta\varepsilon_{\rm ex}}

\newcommand{\T}{\mathcal{T}}

\newcommand{\Lor}{\mathcal{L}}
\newcommand{\vk}{\vec{k}}
\newcommand{\h}{\hslash}

\renewcommand{\Im}{{\rm Im}\,\,}

\newcommand{\dd}{\partial}

\newcommand{\GF}[1]{  \langle\!\langle #1 \rangle\!\rangle  }
\newcommand{\dT}{\Delta T}

\newcommand{\dk}{d^\dagger}

\newcommand{\up}{\uparrow}
\newcommand{\down}{\downarrow}

\newcommand{\beq}{ \begin{equation} } 
\newcommand{\eeq}{ \end{equation} }
\newcommand{\beqa}{\begin{eqnarray}}
\newcommand{\eeqa}{\end{eqnarray}}

\newcommand{\es}{& = &}
\newcommand{\ds}{\displaystyle}

\newcommand{\fig}[1]{Fig.~\ref{fig:#1}}

\hypersetup{
    bookmarks=false,        
    colorlinks=true,        
    linkcolor=blue,        
    citecolor=blue,        
    filecolor=blue,      
    urlcolor=blue
}

\begin{document}

\title{Strong spin Seebeck effect in Kondo T-shaped double quantum dots}
\author{Krzysztof P. W{\'o}jcik}
\email{kpwojcik@amu.edu.pl}
\author{Ireneusz Weymann}
\affiliation{Faculty of Physics, Adam Mickiewicz University, 
			 Umultowska 85, 61-614 Pozna{\'n}, Poland}
\date{\today}

\begin{abstract}

We theoretically investigate the thermoelectric and spin thermoelectric properties 
of a T-shaped double quantum dot strongly coupled to two ferromagnetic leads,
focusing on transport regime where the system exhibits the two-stage Kondo effect.
We study the dependence of the (spin) Seebeck coefficient,
the corresponding power factor and the figure of merit on temperature,
leads' spin polarization and dot level position.
We show that the thermal conductance fulfills a modified Wiedemann-Franz law.
We also demonstrate that the spin thermopower is enhanced at temperatures corresponding to the 
second stage of Kondo screening. Very interestingly, the spin-thermoelectric response 
of the system is found to be highly sensitive to the spin polarization of the leads. In some cases
spin polarization of the order of $1$\% is sufficient for a strong spin Seebeck effect 
to occur. This is explained as a consequence of the interplay between the two-stage
Kondo effect and the exchange field induced in the double quantum dot. All
calculations are performed with the aid of numerical renormalization group technique.

\end{abstract}


\maketitle

\section{Introduction}
\label{sec:intro}

The thermoelectric properties of matter have drawn the attention of physicists since 
the first experiments carried out by Seebeck at the beginning of the 19th century.
While the properties of bulk materials
are already quite well understood \cite{Barnard},
the problem of thermoelectricity in confined nanoscale systems
still contains issues that need further examination, 
although these have been intensively researched since
the famous publications by Hicks and Dresselhaus \cite{Hicks_Dresselhaus_1,Hicks_Dresselhaus_2}.
In particular, thermoelectric and spin-thermoelectric properties of strongly correlated
quantum dot (QD) systems constitute a field of intensive research
\cite{KrawiecPRB06,SwirkowiczPRB09,EspositoEPL09,KuoPRB10,
LiuPRB10,TsaousidouJPCM10,LiuAJP11,MuralidharanPRB12,
SanchezNJP13,KarwackiJPCM13,Hwang15}.
It turns out that the understanding of thermoelectric transport properties
is not only relevant for possible future applications,
but also provides additional information about fundamental
interactions and phenomena at the nanoscale.
One prominent example is undoubtedly the Kondo effect
\cite{Kondo},
which has been attracting the attention of scientists for more than two decades
\cite{goldhaber-gordon_98,cronenwett_98}.
In fact, the Seebeck coefficient for the Kondo quantum dots was not only
reliably calculated \cite{CostiZlatic}, but also measured \cite{Monlenkamp}. 
Moreover, the thermopower was also analyzed for 
double quantum dot (DQD) systems in the isospin Kondo regime,
in which the device was shown to work as
a minimal thermoelectric generator \cite{Donsa}.

\begin{figure}[b]
\centering
\includegraphics[width=0.37\textwidth]{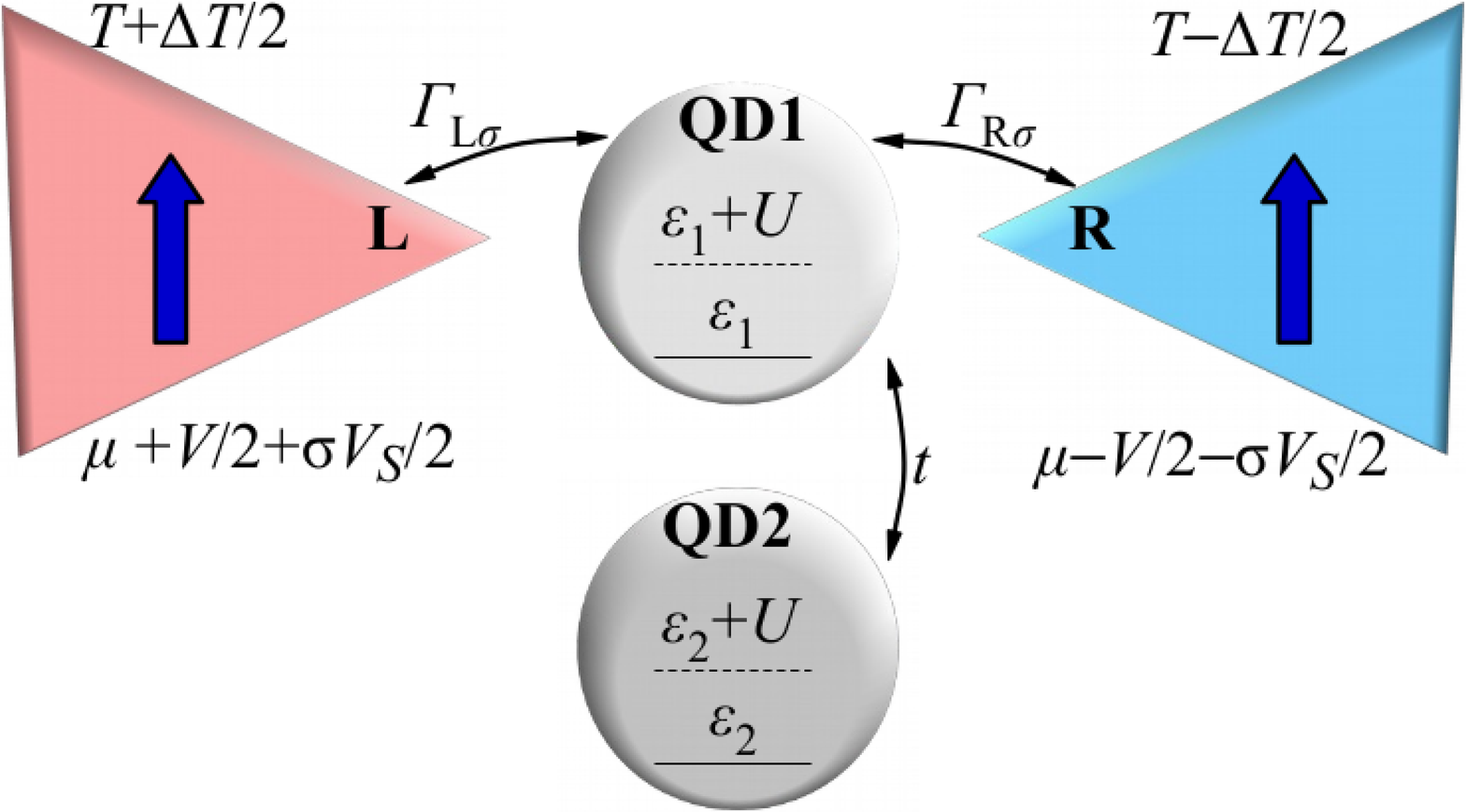}
\caption{Schematic of the system. The left (L) and right (R) leads are coupled to the main
		 quantum dot (QD1) via spin-dependent couplings $\Gamma_{L\s}$ and $\Gamma_{R\s}$.
		 The second quantum dot (QD2) is directly coupled only to QD1, with a matrix
		 element $t$.
		 A small voltage $V$ (correspondingly spin voltage $V_S$) shifts (spin-splits) 
		 otherwise equal chemical potentials $\mu_L = \mu_R = \mu$ symmetrically. 
		 There is also a temperature gradient $\dT$ applied symmetrically
		 to the system.}
\label{fig:system}
\end{figure}

In the presence of magnetic field or when the leads are ferromagnetic,
the thermoelectric response of the system becomes spin polarized
\cite{KrawiecPRB06,SwirkowiczPRB09}.
Spin caloritronic effects of single quantum dots in the Kondo regime
were already studied theoretically
\cite{RejecPRB12,weymannPRB13,weymannSR16}.
Moreover, although spin-resolved thermoelectricity was also
a subject of investigations for DQD systems 
\cite{JB_Trocha2012,KWIW-2qdTermo}, there are still
problems that need further considerations.
In particular, in this paper we analyze the spin caloritronic
properties of a T-shaped DQD coupled to two ferromagnetic leads in the Kondo regime.
The schematic of the considered system is depicted
in \fig{system}. Despite its relative simplicity,
this system hosts a variety of interesting
many-body phenomena.
The screening of subsequent quantum dots gives 
rise to the two-stage Kondo effect, introducing a cryogenic temperature scale $T^*$
associated with the second stage of screening
\cite{PustilnikGlazman,CornagliaGrempel,Zitko2010}. On the other hand,
the dependence of the Kondo temperature $T_K$ and $T^*$
on the DQD level position can lead to Fano-like interference effects 
\cite{Fano,Sasaki,zitkoPRB06,TrochaBarnas,Zitko2010}.
It was shown recently that these phenomena
are reflected in thermoelectric properties of the device \cite{KWIW-2qdTermo}.
Here we extend these studies to more complex, magnetic system.

We note that the influence of magnetism on strongly correlated regime of
T-shaped DQDs was already discussed, but mainly in the context of electrical properties,
such as linear conductance and current spin polarization. The spin-dependent 
Fano antiresonance condition in magnetic field \cite{daSilva} and in
system with ferromagnetic leads \cite{KWIW-spinPol} was predicted. Moreover,
the interplay of the two-stage Kondo screening and the ferromagnets-induced
exchange field was also studied \cite{KWIW-2qd2kondo}. 
The goal of this paper it to investigate the spin thermoelectric properties of the magnetic device.
It is shown that the spin caloritronic coefficients
are strongly affected by the presence of ferromagnetic correlations.
Spin polarization of the order of $1\%$ can already induce a strong spin Seebeck effect
in transport regime where the second stage of screening develops.

We would also like to note that although
direct observation of the spin Seebeck effect was already reported
\cite{Uchida,Flipse}, in quantum dot systems it still remains a challenge.
Therefore, we believe that our results will, on one hand, stimulate further experimental efforts
and, on the other hand, be of assistance in understanding future experimental data.

The paper is organized as follows. In Sec.~\ref{sec:model} the model of 
the device and method used for its solution are explained. 
The relevant energy scales are outlined in Sec.~\ref{sec:scales}.
Main results, concerning the calculated Seebeck and spin Seebeck coefficients 
are presented and discussed in Sec.~\ref{sec:S}.
Finally, Sec.~\ref{sec:conclusions} concludes the paper.

\section{Model and methods}
\label{sec:model}

The considered device consists of two single-level quantum dots
in a T-shaped geometry with the first quantum dot (QD1)
coupled to external ferromagnetic leads and the second dot (QD2)
attached to the first one through the hopping matrix elements $t$,
see \fig{system}. The system can be thus described 
by the following two-impurity Anderson Hamiltonian \cite{SIAM},
$H = H_{\rm DQD} + \sum_r H_r + H_{\rm tun}$.
The first term corresponds to isolated DQD and is given by
\beq
H_{\rm DQD} = \sum_{i\s} \e_{i}  n_{i\s} 
				+ U \sum_i n_{i\up} n_{i\down} 
				+ \sum_{\s} t(\dk_{1\s}d_{2\s} + h.c.)	\, ,
\eeq
where $n_{i\s} = \dk_{i\s}d_{i\s}$ and $\dk_{i\s}$ creates a spin-$\s$ electron 
in dot $i$ with the corresponding energy $\e_{i}$ and
$U$ is the Coulomb correlation parameter in each dot.
The ferromagnetic leads are modeled by free-electron Hamiltonian
$H_r = \sum_{\vk\s} \e_{r\vk\s} n_{r\vk\s}$ ($r=L$ for left and $r=R$
for right lead, $n_{r\vk\s}$ denotes the occupation operator for state 
characterized by momentum $\vk$, spin $\s$ and lead $r$, while $\e_{r\vk\s}$
is the energy of the corresponding level).
The coupling between the first dot and the leads is
described by the tunneling Hamiltonian
$H_{\rm tun} = \sum_{r\vk\s} v_{\vk\s} (\dk_{1\s}c_{r\vk\s} + h.c.)$,
where $c_{r\vk\s}$ is the corresponding annihilation operator
and $v_{\vk\s}$ denotes the respective tunnel matrix element.

We consider the wide-band limit and assume that only $s$-waves couple to the
electrodes. This allows us to write the spin-dependent coupling 
$\Gamma_{r\s} = \pi \rho_{r\s} |v_{\vk\s}|^2$ as a constant 
($\rho_{r\s}$ denotes the normalized spin-resolved density of states 
of lead $r$ at the Fermi level), determined by the leads' spin polarization $p_r$.
For parallel configuration of the magnetizations of the leads,
one then gets $\Gamma_{r\s} = (1 + \s p_r) \Gamma_r/2$, and 
$\Gamma_\s \equiv \Gamma_{L\s}+\Gamma_{R\s} = (1+\s p) \Gamma$,
where $p=(p_L+p_R)/2$ is the effective leads' spin polarization
and we assumed $\Gamma_L = \Gamma_R \equiv \Gamma/2$.
We note that in the antiparallel magnetic configuration,
for left-right symmetric systems, the couplings become spin independent
and the transport properties are similar to those in the nonmagnetic case
with a polarization dependent factor. On the other hand,
when the system is not symmetric, the behavior is the same as in the case
of parallel magnetic configuration with some new coupling strength and effective spin polarization \cite{wojcikJPCM13}.
Therefore, in the following we will consider only the case of parallel magnetic configuration.

Let $I_x$ denote the $x$-current ($x=C$ for charge, $x=S$ for spin, 
$x=Q$ for heat). Using the Boltzmann equation approach and assuming well-defined
Fermi level to be the reference point for energy scale, one can derive the 
linear-response coefficients connecting currents with voltage $V$, spin voltage 
$V_S$ and temperature difference $\dT$ \cite{Barnard} 
\beq
\left( \!\! \begin{array}{c}
	I_C \\
	I_S \\
	I_Q
	\end{array}\!\! \right)
= \sum_\s
\left( \!\! \begin{array}{ccc}
	e^2 L_{0\s} & \s e^2 L_{0\s} & -e L_{1\s}/T \\
	-\s e\frac{\h}{2} L_{0\s} & -e\frac{\h}{2} L_{0\s} 
			& \s \frac{\h}{2} L_{1\s}/T \\
	-e L_{1\s} & -\s e L_{1\s} & \ds L_{2\s}/T
	\end{array} \!\! \right)
\!\!\! \left( \!\! \begin{array}{c}
	V \\
	V_S \\
	\dT
	\end{array} \!\! \right)  ,
\label{I}
\eeq
where $e$ is the absolute value of electron charge,
\beq
L_{n\s} = -\frac{1}{h} \int \omega^n \, \frac{\dd f(\omega)}{\dd \omega} 
				\T_{\s}(\omega) d\omega ,
\label{Ln}
\eeq
$f(\w)$ is the Fermi-Dirac distribution function, and $\T_\s(\w)$ is the spin-resolved
transmission coefficient. Henceforth we will also use notation
$L_n = L_{n\up}+L_{n\down}$ and $M_n = L_{n\up}-L_{n\down}$.

The transport properties can be calculated from Onsager integrals $L_{n\s}$
using Eq.~(\ref{I}). In particular, the electrical and spin conductances
are 
\beqa
G &\equiv& \dd_{V} I_C|_{V_S=0 \atop \dT=0} =  e^2 L_{0}, \\
G_S &\equiv& \dd_{V_S} I_S |_{V=0 \atop \dT=0} = -e\frac{\h}{2} \cdot L_{0}, 
\eeqa
correspondingly, where $\dd_x A |_{y=0}$ denotes partial derivative of $A(x,y)$ with respect to $x$,
while the condition $y=0$ is fulfilled.
Similarly, the heat conductance is given by
\beq
\kappa \equiv \dd_{\dT} I_Q |_{I_C=0 \atop V_S=0}
	= \frac{1}{T} \left( L_2 - \frac{L_1^2}{L_0}\right),
	\label{kappa}
\eeq
where the conditions $I_C=0$ and $V_S=0$ in fact determine $V$ as a function of $\dT$. 
In this paper we focus on Seebeck and spin Seebeck coefficients, denoted correspondingly by
$S$ and $S_S$, 
\beqa
S 	\es G^{-1} \dd_{\dT} I_C \bigg|_{V=0 \atop V_S=0}
	= -\frac{1}{eT} \frac{L_{1}}{L_{0}}, 
	\label{S}\\
S_S	\es G_S^{-1} \dd_{\dT} I_S \bigg|_{V=0 \atop V_S=0}
	= -\frac{2}{\h T} \frac{M_{1}}{L_{0}}.
	\label{SS}
\eeqa
These are related to the (spin) Peltier coefficient
$\Pi = \dd_{I_C} I_Q |_{V_S=0 \atop \dT=0}$ 
($\Pi_S = \dd_{I_S} I_Q |_{V=0 \atop \dT=0}$) 
by $\Pi_{(S)} = S_{(S)} T$. However, we prefer to study 
$S_{(S)}$ instead of $\Pi_{(S)}$, because Seebeck coefficient
better captures caloric properties at low temperatures.
Finally, we can define the (spin) figure of merit 
\beq
Z_{(S)}T = S_{(S)}^2 G_{(S)} T / \kappa,
\label{ZT}
\eeq
which is a measure of thermodynamic efficiency, and the corresponding power factor
\beq
Q_{(S)}	= S_{(S)}^2 G_{(S)} ,
\label{Q}
\eeq
which is related to maximal power of the device
and the performance under the fixed flow conditions \cite{Narducci}.

The transmission coefficient is proportional to the imaginary part of
the QD1's retarded Green function, 
$\T_{\s}(\w) = -\Gamma_\s \Im \GF{\dk_{1\s} | d_{1\s}}^{\rm ret}(\w)$,
which we determine with the aid of the numerical renormalization group (NRG)
method \cite{Wilson,fnrg}, building the full density matrix from states 
discarded during the iteration of the NRG procedure \cite{AndersSchiller,dm-nrg}.
In calculations we use discretization parameter $\Lambda=2$
and keep $2048$ states at each 
iteration. To perform the computations, we assume
flat densities of leads' states within the cutoff $D=2U$
and make a transformation to an even-odd basis \cite{even-odd}.
This leads us to an effective single-channel formulation of the problem, where
for the parallel magnetic configuration the only parameters
corresponding to the conduction bands
are those related to an effective one, namely $\Gamma$, $p$ and $D\equiv 1$.

The NRG method allows us to obtain reliable results in the whole parameter space of the model,
in particular, at finite temperatures. However, NRG forces us to limit our 
considerations to the linear response regime, where a single Fermi level 
can be defined for both leads and, thus, the logarithmic discretization, being
a key ingredient of the procedure, is well defined \cite{Wilson}.

\section{Relevant energy scales}
\label{sec:scales}

\begin{figure}
\centering
\includegraphics[width=0.42\textwidth]{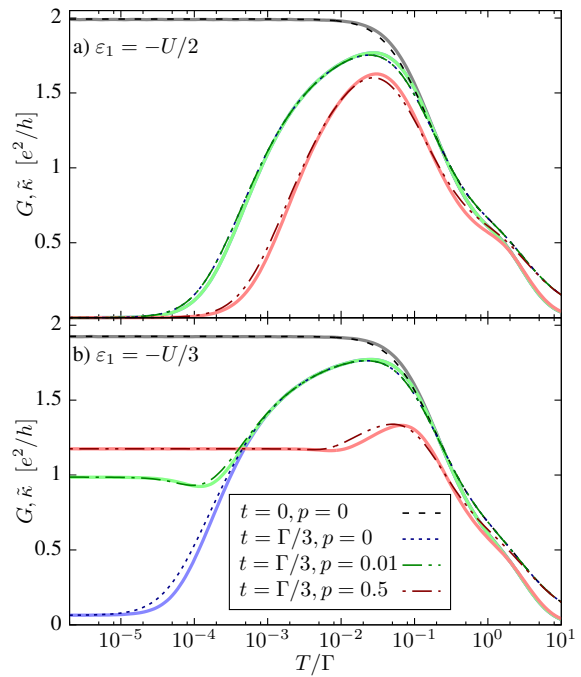}
\caption{The linear conductance $G$ (dashed lines) as a function of
		 temperature $T$ calculated for $\e_2 = -U/2$, $\Gamma=U/5$ and
		 (a) $\e_1 = -U/2$, (b) $\e_1 = -U/3$.
		 Solid lines indicate the rescaled and shifted 
		 thermal conductance, $\tilde{\kappa}\equiv \Lor_0^{-1} 
		 \kappa(\alpha T)/ (\alpha T)$ with $\alpha=2$ (see text for details).
		 In (a) the curves for $p=0$ and $p=0.01$
		 (both for $t=\Gamma/3$) are on top of each other.}
\label{fig:WFlaw}
\end{figure}

The considered device hosts very reach physics at the manifold of energy scales.
This can be seen in particular in the temperature dependence
of the electrical conductance presented in \fig{WFlaw}.
For decoupled second dot ({\it i.e.} for $t=0$), nonmagnetic leads 
($p=0$) and QD1 energy level in the Coulomb valley ($-U \ll \e_1 \ll 0$), at 
temperatures below the Kondo temperature $T_K$, the conduction band electrons 
screen the spin of the electron occupying QD1. This screening
results in an additional resonance in the local density of states
of the first dot at the Fermi level, which 
gives rise to an enhancement of the conductance $G$,
see the curves for $t=0$ and $p=0$ in Figs. \ref{fig:WFlaw}(a) and (b).
In the Kondo regime $G$ can achieve the unitary limit $G=2e^2/h$, if 
the dot is tuned to the point of the particle-hole symmetry (PHS),
$\e = -U/2$, as it is done in \fig{WFlaw}(a). 
The maximal conductance outside the PHS point is slightly smaller, see \fig{WFlaw}(b).
For single quantum dots coupled to ferromagnetic leads,
the Kondo temperature can be estimated from scaling
approach \cite{Haldane,Martinek},
\beq
T_K \approx \sqrt{\frac{\Gamma U}{2}} \; \exp \left[ 
	\frac{\pi \e (\e+U)}{2\Gamma U} \; \frac{{\rm arctanh}(p)}{p} 
	\right] .
\label{TK}
\eeq
Experimentally, the Kondo temperature is typically defined 
as the temperature at which $G=G_{\rm max}/2$.
For parameters assumed in \fig{WFlaw}(a)
in the case of $p=0$ and $t=0$, from the temperature dependence of $G$
we find $T_K \approx 0.32\Gamma$.

The coupling between quantum dots results in the emergence
of another energy scale, $T^*$, which for relatively weak $t \lesssim \Gamma$
is associated with the screening of second dot's spin
by the continuum formed by QD1 and leads.
This screening manifests itself through a decrease of
$G$ for temperatures below $T^*$, see the curves for $t=\Gamma/3$,
$p=0$ in Figs. \ref{fig:WFlaw}(a) and (b). 
At PHS point the conductance drops to $0$
as $G\propto T^2$ \cite{CornagliaGrempel}, 
while outside this point some finite conductance
remains even in the $T=0$ limit, cf. \fig{WFlaw}(b).
The temperature at which the second stage of screening 
takes place can be theoretically estimated from
\cite{CornagliaGrempel,zitkoPRB06}
\beq
T^* = a \; T_K \; e^{-b T_K/J_{\rm eff}},
\label{Tstar}
\eeq
where $J_{\rm eff} = 4Ut^2 / [U^2 - (\e_1-\e_2)^2]$ 
is the effective exchange interaction between the dots
and $a, b$ are numbers of the 
order of $1$. However, similarly to $T_K$,
we estimate $T^*$ numerically from the temperature dependence of $G$,
as the temperature at which the conductance drops to half of its maximum value.
For parameters assumed in \fig{WFlaw}(a), $p=0$ and $t=\Gamma/3$,
$T^* \approx 5.9 \cdot 10^{-4}\Gamma$.

The influence of leads' ferromagnetism on transport properties differs significantly,
depending on the presence or lack of particle-hole symmetry in the system.
At PHS point, $\e_1 = \e_2 = -U/2$, the leads' spin polarization only slightly modifies
$T_K$ and $T^*$, cf. Eqs. (\ref{TK}) and (\ref{Tstar}), 
which also causes some minor change in $G_{\rm max}$, see \fig{WFlaw}(a). 
However, outside the PHS point this influence is much more pronounced,
as can be seen in  \fig{WFlaw}(b).
Even relatively low values of spin polarization (see the curve for $p=0.01$)
block the second stage of Kondo screening,
while the value of $p=0.5$ is sufficient to suppress the Kondo effect completely.
This is caused by the fact that, in the case of ferromagnetic leads,
the renormalization of double quantum dot energy levels
due to the hybridization with electrodes becomes spin-dependent,
which implies that an effective exchange field $\ex$ is induced in DQD.
$\ex$ strongly depends on the DQD level positions,
in particular, $\ex = 0$ at PHS point.
For $t\ll\Gamma$, one can reasonably approximate $\ex$ induced in QD1
by the formula for a single quantum dot \cite{Martinek,KW-ex},
\beq
\ex^{\rm QD1} \approx \frac{2 p\Gamma}{\pi} \log \left| \frac{\e_1}{\e_1+U}\right| 
\label{exQD1}
\eeq
The determination of the exchange field in the second dot, denoted by $\ex^{\rm QD2}$,
is a more subtle problem \cite{KWIW-spinPol,KW-ex}.
Nevertheless, for $t\ll \Gamma$, $\ex^{\rm QD2}$ can be seen as
a consequence of coupling between QD2 and the continuum formed by QD1 
and the leads. The effective spin-dependent coupling
to the second dot $\Gamma_{2\s}$ is then proportional to $t^2/\Gamma_\s = (1-\s p)\Gamma_2$,
with $\Gamma_2 = (\Gamma_{2\uparrow} + \Gamma_{2\downarrow})/2$,
instead of simply $\Gamma_\s$ as in the case of QD1.
Note that $\Gamma_2$ is a function of both $t$ and $p$,
and the effective spin polarization equals $-p$.
Consequently, while the coupling to one of the spin species is larger in the first dot, 
it can be just opposite in the second dot, which implies that 
$\ex^{\rm QD1}$ and $\ex^{\rm QD2}$ can have different signs
\cite{KWIW-spinPol}.
By raising the exchange field, detuning from the PHS point
by changing either $\e_1$ or $\e_2$ will generally suppress
the second stage of the Kondo effect once $|\ex| \gtrsim T^*$
\cite{KWIW-2qd2kondo}.
Moreover, it can also affect the first-stage Kondo effect if $|\ex| \gtrsim T_K$.

In the strong coupling regime and for $p=0$, the modified Wiedemann-Franz 
law was predicted \cite{Wiedemann,CostiZlatic,KWIW-2qdTermo}, which states 
that at $T<T_K$, $\Lor = \kappa(\alpha T)/ [\alpha T G(T)]$ is a constant, 
instead of the Lorentz number $\kappa(T)/ [T G(T)]$. The value of this 
constant equals $\Lor_0 = (\pi^2 /3) \, k_B^2/e^2$, while the scale shift 
$\alpha$ was estimated to be approximately equal $2$. We found the same 
behavior also in the case of finite spin polarization,
although with slightly worse accuracy. This is illustrated
in \fig{WFlaw}, where rescaled and shifted heat conductance $\tilde{\kappa}(T) 
\equiv \Lor_0^{-1} \kappa(2T)/ (2T)$ is plotted as a function 
of $T$ with solid lines. At $T\lesssim T_K$, all curves overlap to good accuracy
with $G(T)$, which implies that the modified Wiedemann-Franz law 
also holds in the case of T-shaped DQDs with ferromagnetic contacts.

\section{Thermopower and spin thermopower}
\label{sec:S}

In this section we present and discuss the results on
the Seebeck and spin Seebeck coefficients.
First, we study their temperature dependence 
and then analyze what happens when the degree of 
leads' spin polarization is varied. Finally, we consider
the dependence of thermoelectric coefficients
on the position of DQD energy levels.

\subsection{Temperature dependence}
\label{sec:Tp}

\begin{figure}[tb]
\centering
\includegraphics[width=0.95\columnwidth]{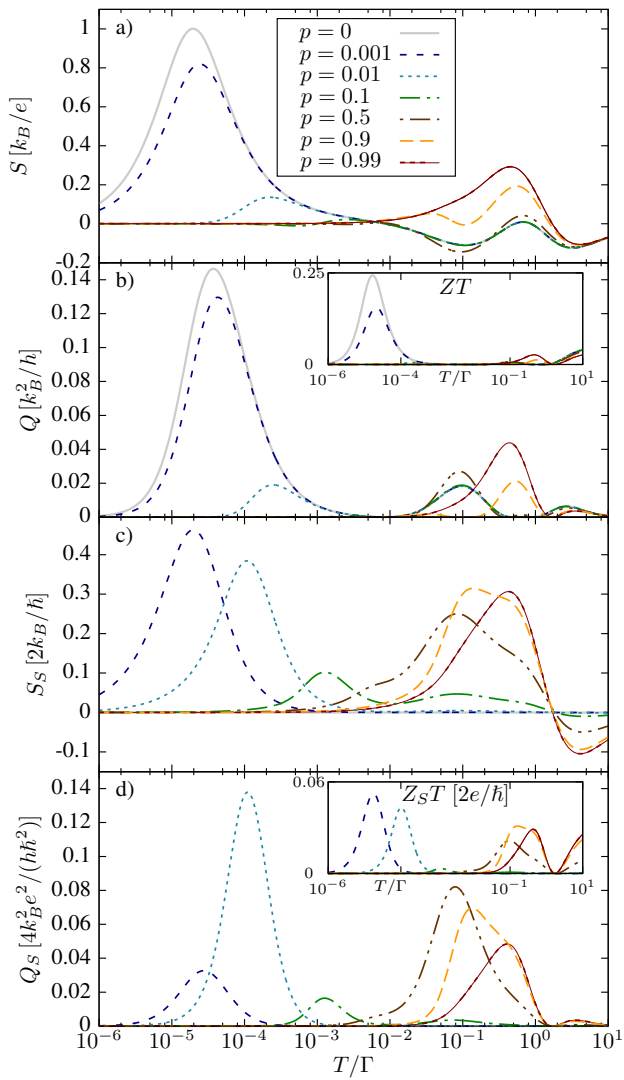}
\caption{Thermopower (a), the corresponding power factor (b)
              and the related spin counterparts (c,d)
              calculated as a function of temperature
              for $\e_1=-U/3$, $\e_2=-U/2$, 
	     $t=\Gamma/3$ and $\Gamma=U/5$,
	     and for different values of spin polarization $p$, as indicated.
	     The insets in (b) and (d) show the 
	     temperature dependence of the
	     corresponding figures of merit, $ZT$ and $Z_ST$.}
\label{fig:S_T_p}
\end{figure}

The full temperature dependence of (spin) thermoelectric 
coefficients is presented in \fig{S_T_p} for different values of leads' spin polarization $p$.
We cover there a wide class of ferromagnetic materials,
starting with nonmagnetic case and ending with half-metals, for which $p\to1$.
This figure was calculated for $\e_1=-U/3$ and $\e_2=-U/2$, 
{\it i.e.} outside the PHS point, since for $\e_1=\e_2=-U/2$,
the thermopower vanishes due to equal contributions from electron and hole processes.
For nonmagnetic systems, the second stage of screening leads
to an enhancement of $S$ at very low temperatures of the order of $T^*$ \cite{KWIW-2qdTermo}.
For finite spin polarization, however, a suppression of the second stage of Kondo effect
by the exchange field occurs, cf. \fig{WFlaw}, 
which suppresses the thermopower peak at $T<T^*$, see \fig{S_T_p}(a).
Clearly, leads' polarization, even as small as $p=0.01$,
is sufficient for the low-temperature peak in $S(T)$ to be strongly suppressed.
This is due to the fact that even very low values of $p$
give rise to finite exchange field, cf. Eq. (\ref{exQD1}),
which for $p=0.01$ can already become larger than $T^*$.
In a similar spirit, larger values of spin polarization
resulting in larger exchange field can affect
thermopower behavior at higher temperatures.
Interestingly, for $T\approx T_K$
one can then observe a more subtle interplay 
between the Kondo correlations and the exchange field.
For $p<0.5$, $S(T)$ exhibits a dip with $S(T)<0$ at $T\approx T_K$,
which is characteristic of the (single-stage) Kondo effect \cite{CostiZlatic}.
On the other hand, with increasing the spin polarization, 
thermopower changes sign and a positive peak appears instead,
see the curves for $p \geq 0.9$ in \fig{S_T_p}(a).
This can be explained as follows. 

\begin{figure}[b!]
\centering
\includegraphics[width=0.42\textwidth]{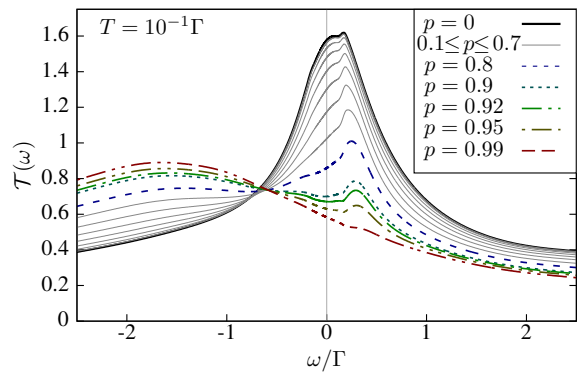}
\caption{The energy dependence of the total transmission coefficient $\T(\w) = \sum_\s\T_{\s}(\w)$
              calculated for different values of spin polarization $p$
              and for parameters corresponding to \fig{S_T_p} with $T=10^{-1}\Gamma$.}
\label{fig:A}
\end{figure}

For $T^*<T<T_K$ and $p=0$ there is a Kondo peak visible
in the total transmission coefficient $\T(\w) = \sum_\s\T_{\s}(\w)$,
as can be seen in \fig{A}, which presents the energy dependence
of $\T(\w)$ for different spin polarization $p$.
Because $\e_1 > -U/2$, the Kondo peak displays some asymmetry
with respect to the Fermi energy ($\w=0$). In fact, finite temperature,
which is slightly below $T_K$, results in a small shift of the maximum to $\w>0$.
Because of that, $\T(\w)$ has a finite slope at $\w=0$,
which is responsible for nonzero thermopower of the device.
For $p\neq 0$ the exchange field appears, which grows with increasing $p$.
Thus, for sufficiently large spin polarization,
$\ex$ can become larger than $T_K$. If this is the case 
the Kondo peak becomes suppressed and split by $2\ex$; see \fig{A}.
Moreover, with increasing the spin polarization,
the levels of DQD become split and
the weight of the transmission coefficient
becomes shifted to negative energies.
This is visible as a gradual enhancement of
the negative-$\w$ Hubbard peak.
For very large spin polarization,
due to the factors $(1\pm p)$,
the majority spin states are mainly responsible for 
the enhanced transmission for $\w<0$.
The above-described behavior results in a sign change of the derivative
of $\T(\w)$ at $\w=0$ with increasing $p$,
which gives rise to the associated sign change of the Seebeck coefficient
visible in \fig{S_T_p}(a).

One could imagine a similar situation for $T\approx T^*$, with a dip in the transmission
coefficient corresponding to the second stage of screening being split by $\ex$.
However, because $\ex$ becomes larger than $T^*$ already for very small values of spin polarization,
e.g. for $p=0.01$ for parameters assumed in \fig{S_T_p}, the 
difference between $(1-p)$ and $(1+p)$ factors
in the spin-resolved transmission coefficient is not significant.
For this reason the relative depth of dips remains approximately 
constant and we do not observe a negative peak at $T\approx T^*$ for any
value of spin polarization considered in \fig{S_T_p}.
However, as presented in Sec.~\ref{sec:pe1},
a sign change of thermopower in the second stage of screening
may occur for $p \approx 0.02$ and is even more pronounced for $\e_1=-U/4$
instead of $\e_1=-U/3$. 

The temperature dependence of the power factor
corresponding to the Seebeck coefficient 
shown in \fig{S_T_p}(a) is presented in \fig{S_T_p}(b).
It exhibits local maxima for temperatures corresponding to peaks (or dips)
visible in $S$, including a small peak at $T\approx U$, associated with thermally excited hole-like
(due to $\e_1<0$) transport.
The inset in \fig{S_T_p}(b) displays the thermoelectric figure of merit $ZT$ as a function of temperature.
It also exhibits all the peaks of $S(T)$, however, the contributions at intermediate temperatures, 
$T^*<T<T_K$, are somewhat suppressed by quite large heat conductance;
see in particular the curve for $p=0.01$ in \fig{S_T_p}(b).

\begin{figure*}[t!]
\centering
\includegraphics[width=1.4\columnwidth]{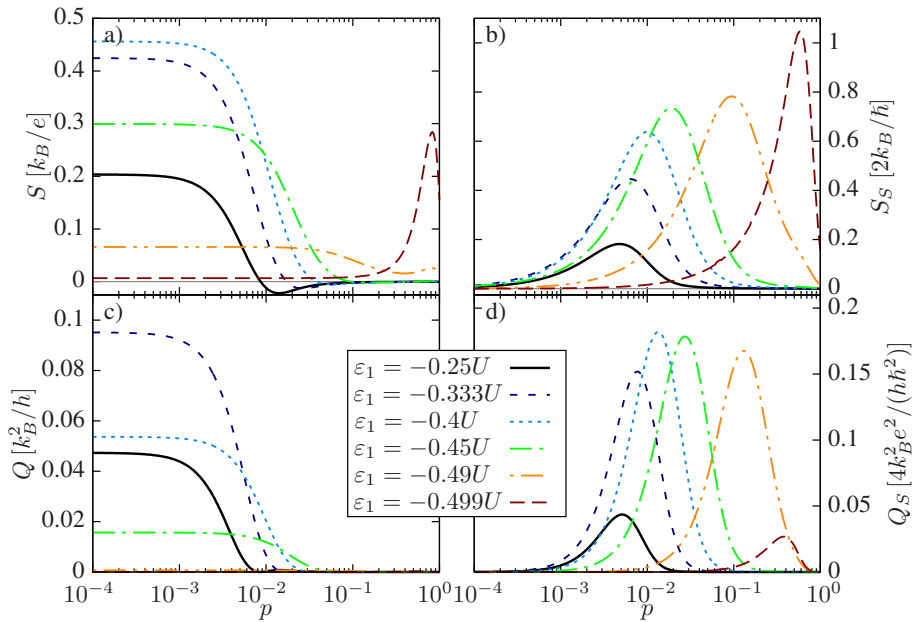}
\caption{The thermopower (a), spin thermopower (b)
             and the corresponding power factors, (c) and (d),
             plotted as a function of spin polarization $p$
             for different values of the level position of the first quantum dot,
             as indicated in the figure. 
             The other parameters are the same as in
             \fig{S_T_p} with $T=10^{-4}\Gamma$.}
\label{fig:S_p_e}
\end{figure*}

We now move to the discussion of spin thermoelectric properties of the considered 
device. A large conventional thermopower present at $T\approx T^*$ gives a hope
that breaking the spin-reversal symmetry by finite leads' spin polarization
will generate a considerable spin thermopower.
However, generation of $S_S$ at such low temperatures
is a matter of a delicate compromise. This is because while
the spin Seebeck coefficient can be generally enhanced by increasing spin polarization,
the second stage of screening and, consequently, the conventional thermopower
become strongly suppressed if $p$ is too large, as already explained in the discussion of \fig{S_T_p}(a).
Nevertheless, as can be inferred from \fig{S_T_p}(c), there are such values
of spin polarization for which the symmetry is sufficiently broken
and a maximum in $S_S(T)$ appears, see the curves for $p=0.001$ and $p=0.01$.
We note that $\h/2\cdot S_S^{\rm max} < e S^{\rm max}$ for $p=0.001$
[$S_{(S)}^{\rm max}$ denotes the maximal value of $S_{(S)}(T)$ for a given value of $p$],
while for $p=0.01$ the opposite inequality holds.
Indeed, in the latter case the spin thermopower
exceeds the conventional thermopower.

With increasing the degree of spin polarization of the leads,
the maximum in $S_S$ moves to larger temperatures
and for $p\geq 0.5$ the spin Seebeck coefficient
exhibits a peak at $T\approx T_K$, see \fig{S_T_p}(c).
Contrary to the case of conventional Seebeck coefficient,
the peak at $T\approx T_K$ has the same sign as the low-temperature peak
for small spin polarization.
This can be surprising, because the peak in $S(T)$
changes sign when $T$ increases from $T \approx T^*$ to $T\approx T_K$, 
cf. \fig{S_T_p}(a).
However, in the case of spin Seebeck coefficient one needs to 
keep in mind that, for assumed parameters, the exchange field
in QD1 is opposite to the exchange field in QD2,
which compensates for this effect. 

One can be surprised that $p$ of the order of one percent
is sufficient to induce a significant spin Seebeck coefficient.
However, it is advisory to recall, that $S_S$ is in fact 
a ratio of a small spin bias $V_S$ and a small temperature gradient $\dT$.
Thus, even the largest value of $S_S$, despite its fundamental aspects,
does not guarantee a practical importance of the result,
if the corresponding power factor $Q_S$ is too small.
For this reason in \fig{S_T_p}(d) we show the temperature dependence of $Q_S$.
It can be clearly seen that the peak in $Q_S(T)$ corresponding to $p=0.001$ is
a few times smaller than the peak corresponding to $p=0.01$.
This remains in agreement with intuition,
that the leads' spin polarization with degree much smaller than $1$\% cannot
induce a significant spin current (although this very small spin current can 
be still much larger then the corresponding charge current).
On the other hand, the peaks visible in $Q_S$ for larger temperatures and
present for strongly polarized leads are also smaller than the peak in the case of $p=0.01$.
This implies that the thermoelectric performance
of the considered device is the best in the regime of 
the second stage of screening.
Nevertheless, the corresponding spin-thermoelectric 
figure of merit $Z_S T$, which is plotted in the inset of \fig{S_T_p}(d),
is not too spectacular, with values only slightly exceeding $0.05 \cdot (2e/\hbar)$.

\subsection{Dependence on the leads' spin polarization}
\label{sec:pe1}

In this section we analyze how the spin thermoelectric properties
depend on the magnitude of the exchange field,
focusing on the second stage of the Kondo effect.
We thus assume the same parameters as in the previous section
and set $T=10^{-4}\Gamma$, which is of the order of $T^*$,
and study the dependence on spin polarization
for different values of QD1 level position.
According to Eq.~(\ref{exQD1}), the exchange field
is linear in $p$ and also in $\e_1$ near the PHS point,
since $\log|\e_1 / (\e_1 + U)| \approx 4(\e_1 + U/2)/U$.
As can be seen in \fig{S_p_e}(a), 
which displays the dependence of $S$ on $p$, 
the Seebeck coefficient in the regime of small
spin polarization is a nonincreasing function of $p$
(note the logarithmic scale for $p$ in the plot).
At sufficiently low spin polarization,
$S$ retains its value for a nonmagnetic system.
However, for any $\e_1$ there is some critical value of $p$,
which we denote $p_c$,
above which the Seebeck coefficient becomes suppressed.
This critical polarization decreases monotonically
with growing the detuning from the PHS point, and is related to some
critical value of the exchange field, $\ex^c\approx T^*$,
overcoming the second stage of the Kondo effect.
As can be seen in \fig{S_p_e}(a), 
the height of $S(p)$ maximum depends on $\e_1$ in a nonmonotonic manner.
This suggest a nontrivial dependence of $S$ on $\e_1$,
which will be explained in the next section.

Moreover, in \fig{S_p_e}(a) we can also notice a small sign change of $S(p)$ for 
$\e_1 = -U/4$ and $\e_1=-U/3$. This is in fact a consequence of the
same phenomenon as that responsible for the sign change
of $S(T)$ for $T \sim T_K$ described in Sec. \ref{sec:Tp}.
The main difference is that the dip in the transmission coefficient,
corresponding to the second stage of the Kondo screening, more 
easily gets smeared, than split. For this reason, the negative peak of 
$S$ is rather small and develops only in a narrow range of parameters, see \fig{S_p_e}(a).

We also note that very close to the PHS point,
one can observe a large peak in $S(p)$,
see the curve for $\e_1=-0.499U$ in \fig{S_p_e}(a).
This result, however, may be considered somewhat artificial.
According to Eq.~(\ref{S}), $S$ is proportional to the ratio 
of $L_1$ and $L_0$. Exactly at the PHS point, $L_1$ is always $0$
while $L_0$ decreases with temperature as $T^2$.
Moreover, $L_0$ is a symmetric function of the detuning from the PHS point,
while $L_1$ is an anti-symmetric function. Thus, for $|\e_1-U/2| \gtrsim T$,
when $L_0$ and $L_1$ are set by detuning, $S$ may reach really large values. 
The role of spin polarization is here to split the transmission coefficient dip and
cause $\mathcal{T}(\w)$ to possess a finite slope at $\w=0$, which additionally enhances 
$L_1$. However, despite large value of $S$, the system does not conduct in this regime
(neither heat, nor current, nor spin), so the result is not really physically interesting.
This is confirmed by the values of $Q$, which is presented in \fig{S_p_e}(c). While
for $\e_1 > -0.49U$, the peaks of $S(p)$ correspond to the peaks of $Q(p)$,
this is not the case for $\e_1 = -0.499U$; $Q$ is not enhanced in the regime of large $p$.

The dependence of the spin Seebeck coefficient $S_S$ on $p$ is presented in \fig{S_p_e}(b). 
It significantly differs from the $p$-dependence of $S$, since $S_S$ vanishes for $p=0$.
In fact, $S_S(p)$ exhibits a peak, whose position varies with $\e_1$ by a few orders of 
magnitude. The height of the peak increases when $\e_1$ approaches the PHS point.
The existence of this peak is a consequence of a balance between the exchange field
and the second stage of screening. If $p$ is small enough, spin-reversal symmetry
is approximately preserved and $S_S \approx 0$. On the other hand,
large values of spin polarization result in strong exchange field,
which destroys the second stage of the Kondo effect, and thus decrease
the spin caloritronic effects.

The exchange field can be also changed by tuning $\e_1$,
which allows for moving the peak in $S_S(p)$ to the desired range of $p$.
The flexibility of the device upon this kind of tuning is reduced by the power factor
corresponding to spin thermoelectric effects, $Q_S$, which is shown in \fig{S_p_e}(d). 
$Q_S$ as a function of $p$ exhibits peaks corresponding to those
present in $S_S(p)$ for all values of $\e_1$ considered.
The height of these peaks is the largest for $-0.49 U < \e_1 < -U/3$ and drops significantly for 
$|\e_1+U/2|<0.01$. This is associated with the suppression of the conductance already 
discussed in the case of conventional Seebeck coefficient.
Moreover, as can be seen in \fig{S_p_e},
for a finite value of spin polarization, there exists such a value of $\e_1$
for which large peaks in $S_S(p)$ and $Q_S(p)$ occur. 

\subsection{Dependence on the position of QD1 energy level}
\label{sec:e1t}

\begin{figure}[tb!]
\centering
\includegraphics[width=0.42\textwidth]{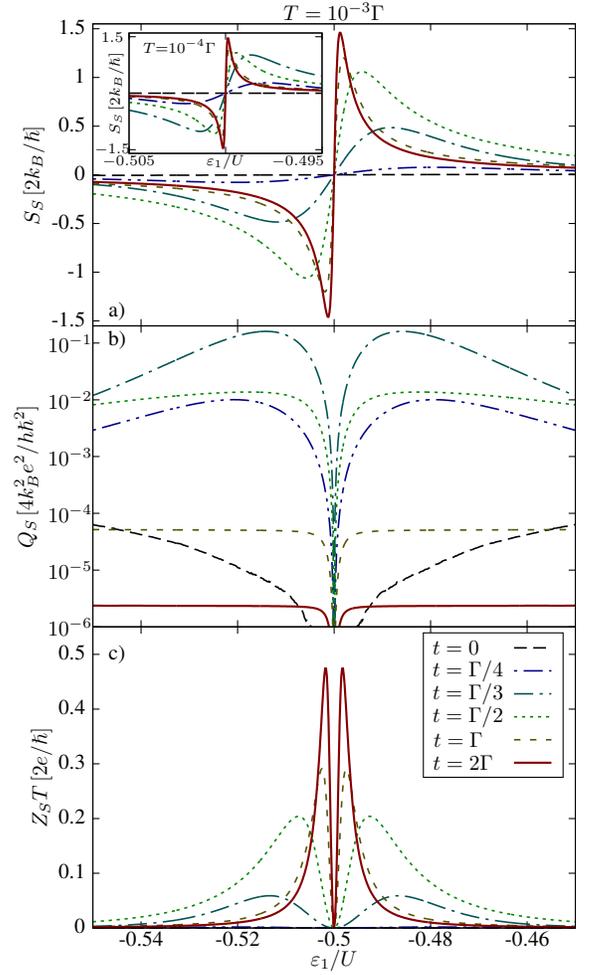}
\caption{The spin thermopower (a), the corresponding power factor (b)
		 and the spin-thermoelectric figure of merit $Z_ST$ (c)
		 as a function of $\e_1$ calculated for
		 different values of hopping between the two dots, as indicated.
		 The parameters are the same as in \fig{S_T_p}
		 with $T=10^{-3}\Gamma$ and $p=0.5$. The inset shows peaks of
		 the spin thermopower at lower $T=10^{-4}\Gamma$.}
\label{fig:S_e_t}
\end{figure}

\begin{figure*}[t!]
\centering
\includegraphics[width=1.4\columnwidth]{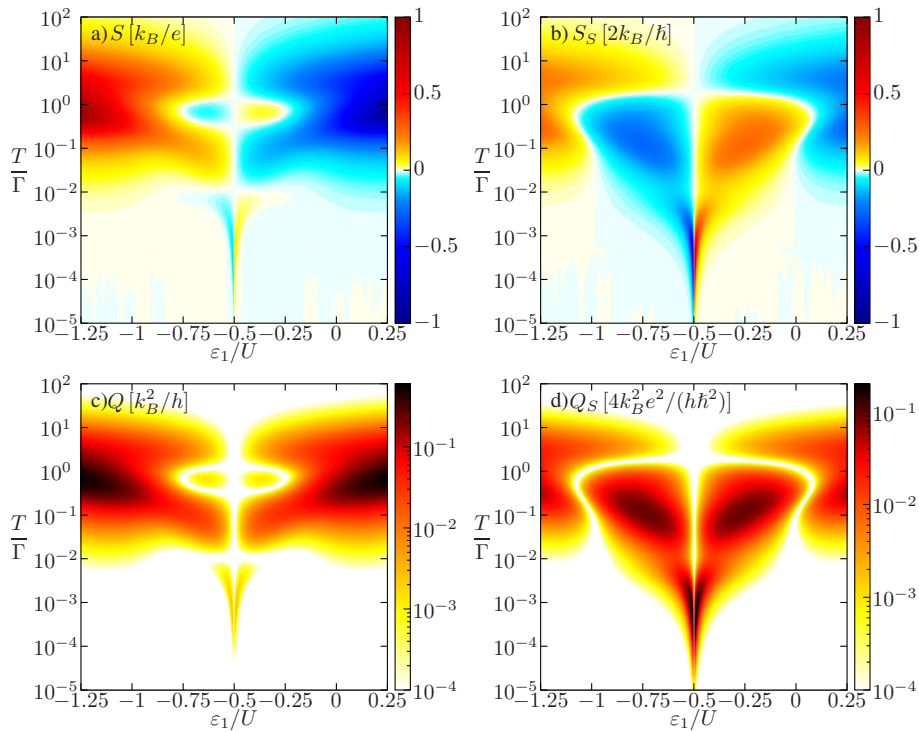}
\caption{The dependence of thermopower (a), spin thermopower (b)
              and the corresponding power factors, (c) and (d), on the first dot level position
              $\e_1$ and temperature $T$.
              The parameters are the same as in \fig{S_T_p} with $p=0.5$.
              Note the logarithmic color scale for the power factors.}
\label{fig:2D}
\end{figure*}

As follows from \fig{S_p_e}, the dependence of $S_S$ on $\e_1$
for large $p$ is quite sharp.
This is related to Fano-like interference,
which occurs between transport paths through
a weakly coupled molecular state of DQD that is a resonant one
and another, strongly coupled state serving as the background \cite{Fano,Sasaki,zitkoPRB06,TrochaBarnas}.
To shed more light on this behavior, 
in \fig{S_e_t} we now plot the full $\e_1$-dependence
of $S_S$ for fixed $p=0.5$ and the other parameters
the same as in \fig{S_p_e}.
In this figure we also study the influence
of different hopping between the dots $t$,
which strongly affects the formation
of molecular states in DQD and,
thus, strongly influences the interference effects.

The dependence of $S_S$ on $\e_1$ calculated at $T=10^{-4}\Gamma$,
{\it i.e.} for temperature corresponding to that used in \fig{S_p_e},
is shown as an inset to \fig{S_e_t}(a).
However, in this case for the considered range of $\e_1$
the spin-thermoelectric power factor $Q_S$ is quite small,
as explained in the previous section (not shown in the plot).
Moreover, it becomes even more suppressed with increasing $t$.
For this reason, the main results shown in \fig{S_e_t} are calculated at larger temperature,
$T=10^{-3}\Gamma$, which is of the order of $T^*$ for $t=\Gamma/3$.
At this temperature, outside the PHS point,
the conductance is not yet fully suppressed due to the second stage of Kondo effect,
and $Q_S$ values are larger, as can be seen in \fig{S_e_t}(b).

At first sight, one can immediately notice a striking qualitative similarity between 
the curves shown in \fig{S_e_t}(a) and those in the inset.
A closer look, however, reveals some differences. 
First of all, the two plots have different scales for the horizontal axis.
It turns out that the sharp interference peaks get broadened with increasing the temperature.
Moreover, the width of those peaks scales approximately linearly with $T$,
while the maximal value of $S_S$ is rather independent of temperature.
We also note that $S_S$ is anti-symmetric around the PHS point,
which is caused by the corresponding sign change of the exchange
field $\ex$ around this point.

The spin-thermoelectric power factor as a function of $\e_1$ is shown in \fig{S_e_t}(b).
One can clearly see that $Q_S$ is optimized for $t=\Gamma/3$ and 
only in this case reaches considerable values.
For smaller values of hopping $t$,
the temperature considered in \fig{S_e_t}(b) is above $T^*$
and $S_S$ is not enhanced.
On the other hand, for larger hoppings,
$T \ll T^*$ and the conductance is generally blocked by the second stage of screening.
Since $Q_S$ must be sufficiently large for any measurement to be 
possible, one should not overestimate the meaning of large spin-thermoelectric 
figure of merit. With this in mind, let us analyze \fig{S_e_t}(c), which presents 
$Z_S T$ as a function of $\e_1$.

As can be seen in the figure, $Z_S T$ exhibits maxima
for such values of $\e_1$ for which $|S_S|$ has peaks.
Due to the square dependence of $Z_ST$ on $S_S$, cf. Eq.~(\ref{ZT}),
the differences in peaks' heights are now more prominent than in the case of $S_S$.
The influence of thermal and electrical conductances compensates
each other. The maximal $Z_S T$ equals $0.5\cdot 2e/\h$, which is quite large, see \fig{S_e_t}(c).
However, it occurs for strong $t$, for which $Q_S$  is rather low
and the measurement is hardly possible.
On the other hand, for $t=\Gamma/3$, corresponding to reasonably 
large $Q_S$, maximal $Z_S T$ remains of the order of $0.1 \cdot 2e/\h$. 

Finally, to make the analysis of (spin) thermoelectric properties of our magnetic
device complete, in \fig{2D} we present the thermopowers and the corresponding power factors 
as a function of temperature and QD1 energy level.
One can see that both $S$ and $S_S$ change sign in the PHS point.
However, $S$ as a function of $T$ exhibits more sign changes
than $S_S(T)$.
The regimes of large Seebeck and spin Seebeck coefficients
can be clearly identified in the figure.
While for $S$ and $Q$ the largest values 
are obtained at relatively high $T$
and large detunings from the PHS point,
$S_S$ and $Q_S$ are maximized for temperatures of the order of
$T^*$ and close to (but not at) the PHS point. 

\section{Conclusions}
\label{sec:conclusions}

We have analyzed the thermoelectric and spin-thermoelectric properties of the 
double quantum dot in a T-shaped configuration, coupled to two leads magnetized in parallel.
The calculations were performed in the linear response regime with the aid of the NRG
and we focused on the parameter regime where the system exhibits the two-stage Kondo effect.
We determined the full temperature dependence of the (spin) Seebeck coefficient,
together with the corresponding power factor and figure of merit.
We also studied the dependence of the spin caloritronic properties 
on the degree of spin polarization of the leads, dot level detuning
and the strength of hopping between the dots.
It was demonstrated that the thermal conductance
fulfills the modified Wiedemann-Franz law
found previously for nonmagnetic systems.
In addition, we showed that the spin Seebeck coefficient
can be strongly enhanced in the regime corresponding 
to the second stage of the Kondo effect.
This enhancement is very sensitive to the value of leads' spin polarization.
Moreover, it can be tuned by changing the DQD parameters,
such as level position and hopping between the dots.
We also showed that in order to keep the power factor
at an experimentally relevant level,
one needs to set the temperature of the order of $T^*$.
Since $T^*$ strongly depends on $t$, this effect can be tuned by
changing the hopping between the dots and the temperature.

We would also like to emphasize that the spin thermoelectric properties 
of the considered device are very sensitive to the spin polarization of the leads,
and even small values of $p$ (of the order of $1\%$) can induce large spin Seebeck effect.
Such a value of spin polarization may be a consequence of 
current-induced spin accumulation even for very small driving currents.
It can also occur in the case of the anti-parallel configuration of leads' 
magnetizations for two asymmetrically coupled electrodes.
Then, even very small coupling asymmetry changes the effective spin polarization
from $0$ to a finite value of $p = (\Gamma_L-\Gamma_R)/\Gamma$.
It therefore seems quite realistic to expect $p \gtrsim 0.01$ in an experiment,
which will cause the conventional Seebeck effect
to be strongly suppressed (compared to nonmagnetic case)
and the spin Seebeck effect to be present and even possibly strong.
All these implies that the effects studied in this paper
may be also relevant for a system,
in which one would not expect them to appear.

\acknowledgments
This work was supported by the National Science Centre in Poland
through Grant No. DEC-2012/04/A/ST3/00372.

\end{document}